\documentstyle[12pt]{article}
\begin{document}
\title
{\bf
Block spins and Chirality in Heisenberg model on Kagome
and triangular lattices}
\author{
V. Subrahmanyam\cr
International Center for Theoretical Physics, P.O. Box 586\cr
34100 Trieste, Italy\cr }
\maketitle
\begin{abstract}
The Spin-${1\over2}$ Heisenberg Model (HM) is investigated using
a block-spin renormalization approach on Kagome and triangular lattices.
In both cases, after coarse graining the triangles on original lattice and
truncation of the Hilbert space to the triangular ground state subspace,
HM reduces to an effective model on a triangular lattice in
terms of the triangular-block degrees of freedom $viz.$
the spin and the chirality quantum numbers. The chirality part of the effective
Hamiltonian captures the essential difference between the two lattices.
It is seen that simple eigenstates can be constructed for the effective model
whose energies serve as upper bounds on the exact ground state energy of HM,
and chiral-ordered variational states have high energies compared to the other
variational states.

\end{abstract}
\vskip 1in

\newpage
\hoffset = -1truecm
\parindent=0.5in
\baselineskip=0.8truecm
Antiferromagnetism in frustrated lattice systems is
known to display some physically striking features,
for instance a finite ground state entropy\cite{kn:Dom60}. Heisenberg spin
systems on
frustrated lattices has been investigated extensively using diverse methods
\cite{kn:And73, kn:Hus88, kn:Sac90},
and ground states with novel structures have been proposed\cite{kn:Kal87,
kn:Cha90}.
Many authors have addressed the question of a chiral
long range order, especially after a suggestion of its possible connection
with the high-$T_c$ superconductors\cite{kn:Kal87,kn:Wen90,kn:Bas89}. In this
letter we investigate Heisenberg spin systems
using a block-spin renormalization scheme on lattices whose basic unit is a
triangle, which is the smallest system with frustration and a chirality.
After coarse graining
the triangles on the original lattice we derive an effective Hamiltonian which
explicitly shows the form of the chiral interactions between the blocks.

We consider a spin-1/2 Heisenberg antiferromanget on triangular and Kagome
lattices. The Hamiltonian
is given by
\begin{equation}
{\cal H} =  J\sum_{ij} {\vec s_i}.{\vec s_j},
\label{eq:ham}
\end{equation}
where the sum is over all the bonds on the Kagome lattice (KAF)
or triangular lattice (TAF), ${\vec s}_i$ is a spin-1/2 operator at site $i$,
and $J$ is the exchange interaction strength. The classical states (which
corresponds to the limit $s\rightarrow\infty$) are known
for both the lattices, where the angle between any pair of spins is
$2\pi/3$, and all the spins on a triangle are coplanar. In contrast, in the
spin-1/2 ground state on a triangle the spins are not on a plane,
and the chirality is a measure of this noncoplanarity.

We are interested in developing a systematic coarse graining
procedure capturing the essentials of frustration, and finding explicitly the
block-spin effective interaction.  Also we can see
the difference between the effective triangular interaction on the Kagome
and the triangular lattices, as the amount of frustration is different in the
two cases. We start with
blocking the original lattice into independent triangular blocks. Using the
eigenfunctions of the triangles we derive a block-spin
Hamiltonian in terms of the block degrees of freedom, namely the total spin
and
the chirality. This procedure will be an exact transformation if all the
8 states
per triangular block are kept. However we will effect a truncation of the
Hilbert space
at the block level by keeping only half the number of states, $i.e.$
restricting
the Hilbert space to the block ground state subspace, as is
explained below. This procedure is equivalent to doing a perturbation
theory on the inter-block interaction. A similar scheme has been implemented
for the Heisenberg model on $C_{60}$ recently\cite{kn:Sub93}.
For KAF we block all the triangles standing upright (as shown in Fig.1), and
in the case of TAF we use the blocking scheme used by Niemeijer and
van Leeuven for the Ising problem\cite{kn:Nei73} (here one third
of the triangles standing upright are blocked). In both cases the interactions
between the blocks is mediated by the inverted triangles ($i.e.$ triangles
standing on one vertex).
The new inter-block Hamiltonian after coarse graining is defined on
a triangular lattice with the total number of sites a third of the original
lattice.

The Heisenberg model on a triangle has two four-fold degenerate energy levels
with the total spin $S=1/2,$ and $3/2$. The ground state with $S=1/2$ has
an energy $-{3\over4}J$,
and the $S=3/2$ excited states are $1.5 J$ above the ground state, which we
will drop by halving the Hilbert space.
A two-fold Kramers degeneracy in the ground state is implied as we have an odd
number of spins. The extra
degeneracy comes from the chirality of the triangle. The chirality operator
for a triangle
is defined\cite{kn:Wen90} through $\chi={2\over \sqrt{3}}{\vec s}_1.{
\vec s_2} \times {\vec s_3}$. We have inserted a numerical factor along
with the box product of spins so as to make the chirality operator a spin-1/2
operator. The abvoe operator can be chosen to be the $z-$component of the
chiratlity operator ${\vec \chi}$.
It is not usual in the literature to treat the
chirality as a spin-half operator. However, within the ground state subspace
for a triangle the two-fold degenerate ground state in both the spin sectors
can be thought of as a chiral spin-1/2 system\cite{kn:Sub94}. Inter-block
interactions cause transitions between the two chiral states,  thus
the lowering and
raising chiral operators naturally arise, which we write in terms of
the of the original spin operators later.
To be consistent with labeling as the chirality changes
sign under odd permutation of spin labels, chirality always refers to
the chirality of a triangle standing upright
with the first spin at the vertex and the second spin at the left corner of the
base. Let us label
the four ground states using the two block quantum numbers $S^z=\pm 1/2$, and
the chirality $\chi^z=\pm 1/2$. The state with $S^z=1/2,\chi^z=1/2$ denoted by
$|++>$ is given in the $s^z$ basis as
\begin{equation}
|++>={1\over \sqrt{3}} |\uparrow\uparrow\downarrow> +
{\omega\over \sqrt{3}} |\uparrow\downarrow\uparrow> +
{\omega^2\over \sqrt{3}} |\downarrow\uparrow\uparrow>,
\label{eq:tgs}
\end{equation}
where $\omega$ is the cube root of unity. The other states can be generated by
interchanging $\omega$ and $\omega^2$ (this changes the chirality from + to
- with the $S^z$ fixed), and by operating with the total spin-lowering operator
(this changes $S^z$ from 1/2 to -1/2 with the chirality fixed).
Let us construct the chiral raising operator
its hermitian conjugate by
defining $\chi^+|+>=0,$ and $\chi^+|-> = |+>$ in both the spin sectors.
It should be noted that there is local degree of freedom for each triangle,
that is we can choose any arbitrary linear combinations of the two
chiral states, as the Hamiltonian does not have explicit chirality terms.
This can be used to advantage in the effective Hamiltonian we derive below.

We would like to write an effective Hamiltonian between the blocks in terms of
the total block spin ${\vec S}_i$ and the chirality ${\vec \chi}_i$ of the
blocks.
Since the original Hamiltonian has only pair-wise interactions, the problem
reduces to a two-triangle problem (with a 16-dimensional Hilbert space) which
can be done analytically.
That is we express $<\psi_l|{\vec s_i}(T1).
{\vec s}_j(T2)|\psi_m>$ as an operator, where $l$ and $m$ are labels
on the wavefunctions of two-block system, $viz.$
direct products of eigenfunctions of two triangles $T1$ and $T2$, which are
connected through the
spins $s_i$ and $s_j$. To accomplish this we need to know the action of
original spin operators on the triangle eigenfunctions, $i.e$ for instance
for a given triangle we have
$$s^z_1|++>= 1/2|++> - {\omega^2\over \sqrt{3}}|\downarrow\uparrow\uparrow>,$$
$$s^z_1|+->= 1/2|+-> - {\omega\over \sqrt{3}}|\downarrow\uparrow\uparrow>,$$
and similar relations involving the other states and operators.
We can anticipate that the effective interaction between
the block spins will be isotropic, as we have not broken the rotational
symmetry in spin space by our blocking procedure. The spin part of the operator
factors, and we are left with a four-state problem. We can explicitly carry
out the evaluation of the above matrix elements and write
the effective interaction
between two triangles as a product of spin and chiral interactions.
The details of the calculation will not be given here. The effective
Hamiltonian is given as
\begin{equation}
{\cal H}_{eff}= {16\over 9}J \sum {\vec S_i . \vec S_j} (H_{ij}+U_{ij}+
D_{ij})
\label{eq:eff}
\end{equation}
where the operators $H,D,U$ are nonzero on horizontal, upward and downward
bonds respectively on a triangular lattice as shown in Fig.2. The explicit
form of these bond operators will be given below.
It is interesting to see that bonds now carry arrows as shown,
and all the bonds in one direction have the arrow in the same
direction. Inside a triangle the arrow is in one direction only. If we assign
a new chiral variable to the direction of the arrow for a given triangle, two
neighboring touching triangles have opposite chirality. Let us define the
operators $T^A,T^B,T^C$ for
every block in terms of the raising and lowering chiral operators through
$ T^A =\chi^++\chi^-/2$,
$T^B =\omega\chi^++\omega^2\chi^-/2$,
$T^C =\omega^2\chi^++\omega\chi^-/2$.
In terms of these operators, the bond interactions in Eq.\ref{eq:eff} are
given as
$$H_{ij}=({1\over4}-T^A_i)({1\over4}-T^C_j),
{}~U_{ij}=({1\over4}-T^B_i)({1\over4}-T^A_j),
{}~D_{ij}=({1\over4}-T^C_i)({1\over4}-T^B_j)~{\rm for KAF},
$$
$$H_{ij}=({1\over2}+T^A_i)({1\over4}-T^A_j),
{}~U_{ij}=({1\over2}+T^B_i)({1\over4}-T^B_j),
{}~D_{ij}=({1\over2}+T^C_i)({1\over4}-T^C_j)~{\rm for TAF}.
$$
As we can see from the above the chirality part of the effective Hamiltonian
is very different for the two lattices. Also it is peculiar that there are no
terms involving $\chi^z$ operators, and the interaction in x and y directions
is anistropic and not very intuitive. In both cases the bonds are directional,
though the way the bonds become directional in the two cases is different.
For the Kagome lattice the orientation of a given block with respect its
nearest neighbor blocks gives the directionality to the effective bond
strength.
There is an additional source of directionality for the triangular lattice,
there are two bonds from a single site of one block going to
to two different sites of another block. It is interesting to see, though
the effective model on a triangular lattice is also frustrated, if the
the block spin and the chirality can conspire so as
to cancel some of the frustration effects, which we investigate later.

The $T$ operators are
linearly dependent $T^A+T^B+T^C=0$, and we will see below that these are
related
to the permutation operators of the triangular block. The eigenfunctions of
the operator $T^A$ in $\chi^z$ basis are $|+>\pm|->/\sqrt{2}$. The
operators $T^B$ and $T^C$ have expectation values $\pm1/4$ in eigenstates of
$T^A$ with eigenvalues $\mp1/2$.
By choosing a given
linear combination of the two-fold degenerate chiral states for a given block
we can monitor the bond strength for the block spins! Thus we have a variety
of variational wavefunctions that can be constructed for the effective model
with the bond strengths transfered to the desired singlet valence bonds.
Let i'th block be in
state $\phi_i=a_i|+>+b_i|->$, where the kets denote the chiral states. Then
the $T$ operators have the expectation values  $T^A=2c_i,T^B=-c_i+{\sqrt{3}
\over2}
d_i,T^C=-c_i-{\sqrt{3}\over2}d_i$, where $c_i={\it Re}(a_i^*b_i)$ and $d_i={\it
Im}
(a_i^*b_i)$.
Now we have variational parameters $a_i$ explicitly in the Hamiltonian itself.
which can be chosen to minimize the energy.
Let us try a simple choice $\phi_i$
such that $T^A_i=-1/2,T^B_i=1/4,T^C_i=1/4$, and
$T^A_i=1/4,T^B_i=1/4,T^C_i=-1/2$ at alternate sites for KAF. This makes all
the bond strengths zero except half the horizontal bonds
(with bond strength $J$).
The ground state is just a singlet spin state on these bonds. This gives
us a bound on the ground state energy per site for KAF, $E_{gs}(KAF)\le-3/8J$.
We expect that a careful choice of $\phi_i$ one can get a better bound.
For the triangular lattice problem a choice with $T^A=1/2$ and $T^A=-1/2$ at
alternate sites along with spin singlets on horizontal bonds of strength
${4\over3} J$ yields a bound $E_{gs}\le -{5\over12} J$.

Let us consider the effective Hamiltonian of KAF and TAF on a triangle. Since
the effective bond interaction is directional (see Fig.2), we have to examine
two types
of triangles, $viz.$ a triangle standing upright and a triangle standing on
a vertex. For the KAF effective model the maximum strength of any bond is
$J$ in any variational state. In contrast for TAF we can have bonds of strength
${4\over3}J$. This means valence bond state has an energy of -3/4 and -1 for
KAF
and TAF respectively in units of $J$, which is also the energy for a bare
Heisenberg Hamiltonian on a triangle. But we can do better than this by taking
advantage of the more number of states we have in the case of the effective
model (64 states). In fact it is borne out by the exact diagonalization we
have done numerically. The ground state is in $S=1/2$ sector with four-fold
degeneracy, implying a new chirality. The energy is -0.987 for an upright
triangle (more than 30\% lower than a valence bond state) and
-3/4 for the upside down triangle for KAF, where as for TAF both
triangles have an energy -1. This gives an indication that for KAF the
frustration is reduced considerably at the first level of blocking, and
presumably it should get better on further blocking (for instance blocking the
less-frustrated triangles at the second level). Let us write the ground state
energy per site of KAF as $ E_{gs}(KAF)/J=\sum_{i=1}^{\infty} {x_i/3^i}$,
where $x_i$ is the energy of a triangular unit at $i'$th blocking. We have
the first two terms of the series, $x_1=-3/4$, and $x_2=-0.987$. Let $y_i=
x_i/x_{i-1}$ be the ratio of the energies of the basic unit at successive
blockings. If we assume that after each blocking $y_i$ does not decrease,
$i.e.$ $y_i \le y_2=1.316$, which
is supported by the fact that the frustration effects are not as strong after
blocking as discussed above, we can get a bound on the ground state energy,
$E_{gs}\le x_1/ (3-y_2)=-0.445$. We need at least a few more
$x_i$ (which could be computed numerically) to support this estimate, but
indication is in favor of this. Similarly for TAF we have $x_1=3/4, x_2=1$,
$y_2=4/3$, this gives us a bound $E_{gs}(TAF)\le -9/20$.
In this case, it may be a
good idea to try
variational valence bond wavefunctions with the effective Hamiltonian.
This is because of the lack of indication of frustration growing less at this
first level of blocking and the
Hamiltonian is difficult to carry out the next level blocking analytically.
The valence bond variational states can exploit the ablility of turning off
or on of bonds, and can give better upperbounds.

Now we will rewrite the effective Hamiltonian in a more transparent form.
Firstly, we note that the inter-block interactions are causing transitions
between the two chiral states of a block, as it is clear from the appearance
of $\chi^+$ operators. The chirality of a block is changed by a
permutation of the spin labels. This implies one can explicitly construct the
chirality operators using the permutation operators.
Let us define permutation operator $P1$,
which permutes the spins 123 to 132 of a given block,
and similarly $P_2$, and $P_3$ exchange spins 1 and 3, and 1 and 2
respectively.
The $P$ operators can be written in terms of the spin operators\cite{
kn:Dom60}, for instance $P_1=2({\vec s}_2.{\vec s}_3+1/4)$.
The action of the permutation operators on the chiral states is seen
explicitly,
$P_1|+>=\omega|->$, and $P_1|->=\omega^2|+>$.
Let us denote by
$e_i$ the operator of a bond
opposite to site $i$ in a block. We have $<e_1+e_2+e_3>=-3/4$,
and $-3/4 \le <e_i>\le 1/4$.
The chirality operator $\chi^z$ is defined as a commutator of $P$
operators through
\begin{equation}
\chi^z\equiv {i\over 2\sqrt{3}} [P_2,P_1]={2i\over \sqrt{3}}(e_2e_1-e_1e_2),
\end{equation}
and the operator $\chi^x$ is just half of $P_3$. It is interesting to note that
a similar construction can be used to
construct the chirality operators in terms of the spin operators directly
even in a general case of more than three spins\cite{kn:Sub94}, in contrast
with the usual  practice of defining $\chi^z$ in terms of fermion
operators\cite{ kn:Wen90}.

It is easy to check that the $T$
operators appearing in the effective Hamiltonian are related to the permutation
operators through $T^A=P_3/2,T^B=P_2/2$, and $T^C=P_1/2$.
The effective Hamiltonian has a very simple form in terms of the bond
operators.
For instance two blocks $l$ and
$m$ on Kagome lattice, with block spins $\vec S_l$ and $\vec S_m$,  connected
at sites $i(l)$ and $j(m)$ have the effective interaction
$$
{16 \over 9}J {\vec S}_l.{\vec S}_m e_i(l)e_j(m).
$$
This has a physical significance, in terms of a valence bond trial
wavefunction,
that the block spins prefer that the bonds opposite to the connecting
sites to be in singlet so that the block spins can form a singlet too.
The frustration of the original lattice translates into frustration for the
block spin bonds. For TAF the scenario is
different, as a pair of blocks standing upright are connected by
two bonds of the original lattice, between the vertex of $l'$th block
(vertex site $i(l)$) and the base of $m'$th block with (vertex site i(m)).
The interaction between the block spins is $-16/9J {\vec S}_l.{\vec S}_m
e_i(l)(3/4+e_i(m))$.

With the effective bond strengths expressed in terms of
the original bond operators, we can now investigate chiral-ordered states of
the original lattice.
Let us try a trial wavefunction $|\psi>$ with chiral ordering on the original
lattice, which implies for each block we choose one of the chiral states.
In any of the four chiral states of a given block,
all the bond operators have an expectation value $e_i=-1/4$. This means
that the block spins interact with a strength of 2/9J for TAF, and 1/9J
for KAF for this trial state. This gives us a bound $E_{TAF}(\psi)\le-{1\over
4}+
{2\over27}E_{TAF}$, implying $E_{TAF}(\psi) \le -0.27$, which is significantly
larger than the energy of the valence bond state we discussed above.
Similarly for the case of
KAF, one can see that chiral-ordered variational states have energies well
above the other variational states.
However one can try other variational states, particularly the RVB-type states
are very convenient to work in this respect. If a triangle has a singlet
valence
bond it implies that one of the $e_i$  is equal to -3/4 and the others vanish.
That is four out of the six bonds incident on a
particular site on the effective
lattice have exchange strength zero. And
the expectation values are calculated straightforwardly, as the effective
model decomposes into cluster hamiltonians. An investigation of
various variational trial states on the effective model is in progress.

We would like to add that this
procedure is also equivalent to using the real-space renormalization
prescription in terms of the density matrix of the block given by
White\cite{kn:Whi92}. This is because the density matrix does not mix states
with different spin, though it mixes the excited states with the same spin
as that of the ground state. However, in the case of a bigger block,
for instance a pentagon, with spin-1/2 excited states the two procedure are
not equivalent.
We would also like to point out that one can in fact include all the 8 states
of a triangle. The effective model would be similar, except that it is
difficult
to work with due to the appearance of spin-3/2 projection operators.

In summary, we have established an effective Hamiltonian in terms of block
spins and chirality for both Kagome and triangular lattices, using a block-spin
renormalization approach. The frustration
effects are seen to reduce considerably for the Kagome lattice. The effective
model has a simple physical way of understanding in terms of the permutation
and bond operators of the triangular blocks of the original lattices. The
bond operator expectation values of the blocks can be effectively used in
variational valence bond wavefunctions.

\noindent{\bf Acknowledgement}

It is a pleasure to thank M. Barma and G. Baskaran for useful suggestions
and P. M. Gade and A. M. Sengupta for discussions.

\newpage
\center{\bf FIGURE CAPTIONS}

\noindent{{\bf Figure 1:} A part of the original (A) Kagome (B) triangular
lattices. The blocked triangles are indicated with dots. The effective
model is defined on a new triangular lattice with
the number of sites reduced to a third of the original lattice.}

\noindent{{\bf Figure 2:} The effective block-spin interactions for (A) Kagome
(B) triangular lattices. The arrows carry the information of the relative
orientation of the blocks on the orignial lattices. All the horizontal bonds
have arrows in one direction only, and similarly the other bonds.}
\end{document}